# Incorporating social opinion in the evolution of an epidemic spread

Alejandro Carballosa, Mariamo Mussa-Juane and Alberto P. Muñuzuri*

Institute CRETUS. Group of Nonlinear Physics. Fac. Physics. University of Santiago de Compostela. 15782 Santiago de Compostela, Spain

* alberto.perez.munuzuri@usc.es

Attempts to control the epidemic spread of COVID19 in the different countries often involve imposing restrictions to the mobility of citizens. Recent examples demonstrate that the effectiveness of these policies strongly depends on the willingness of the population to adhere them. And this is a parameter that it is difficult to measure and control. We demonstrate in this manuscript a systematic way to check the 'mood' of a society and a way to incorporate it into dynamical models of epidemic propagation. We exemplify the process considering the case of Spain although the results and methodology can be directly extrapolated to other countries.

## 1. Introduction

Both the amount of interactions that an infected individual carries out while being sick and the reachability that this individual has within its network of human mobility have a key role on the propagation of highly contagious diseases. If we picture the population of a given city as a giant network of daily interactions, we would surely find highly clustered regions of interconnected nodes representing families, coworkers and circles of friends, but also several nodes that interconnect these different clustered regions acting as bridges within the network, representing simple random encounters around the city or perhaps people working at customer-oriented jobs. It has been shown that the most effective way to control the virulent spread of a disease is to break down the connectivity of these networks of interactions, by means of imposing social distancing and isolation measures to the population [1]. For these policies to succeed however, it is needed that the majority of the population adheres willingly to them since frequently these contention measures are not mandatory and significant parts of the population exploit some of the policies gaps or even ignore them completely. In diseases with a high basic reproduction number, i.e., the expected number of new cases directly generated by one infected case, such is the case of COVID19, these individuals represent an important risk to control the epidemic as they actually conform the main core of exposed individuals during quarantining policies. In case of getting infected, they can easily spread the disease to their nearest connections in their limited but ongoing everyday interactions, reducing the effectiveness of the social distancing constrains and helping on the propagation of the virus. Measures of containment and estimating the degree of adhesion to these policies are especially important for diseases where there can be individuals that propagate the virus to a higher number of individuals than the average infected. These are the so-called super-spreaders [2, 3] and are present in SARS-like diseases such as the COVID19. Recently, a class of super-spreaders was successfully incorporated in mathematical models [4].

Regarding the usual epidemiological models based on compartments of populations, a viable option is to introduce a new compartment to account for confined population [5]. Again, this approach would depend on the adherence of the population to the confinement policies, and taking into account the rogue individuals that bypass the confinement measures, it is important to accurately characterize the infection curves and the prediction of short-term new cases of the disease, since they can be responsible of a dramatic spread. Here, we propose a method that quantitatively measures the state of the public opinion and the degree of adhesion to an external given policy. Then, we incorporate it into a basic epidemic model to illustrate the effect of changes in the social network structure in the evolution of the epidemic. The process is as follows. We reconstruct a network describing the social situation of the Spanish society at a given time based on data from social media. This network is like a radiography of the social interactions of the population considered. Then, a simple opinion model is incorporated to such a network that allows us to extract a probability distribution of how likely the society is to follow new opinions (or political directions) introduced in the net. This probability distribution is later included in a simple epidemic model computed along with different complex mobility networks where the virus is allowed to spread. The framework of mobility networks allows the explicit simulation of entire populations down to the scale of single individuals, modelling the structure of human interactions, mobility and contact patterns. These features make them a promising tool to study an epidemic spread (see [6] for a review), especially if we are interested in controlling the disease by means of altering the interaction patterns of individuals. At this point, we must highlight the difference between the two networks considered: one is collected from real data from social media and it is used to feel the mood of the collective society, while the other is completely in-silico and proposed as a first approximation to the physical mobility of a population.

The study case considered to exemplify our results considers the situation in Spain. This country was hard-hit by the pandemic with a high death-toll and the government reacted imposing a severe control of the population mobility that it is still partially active. The policy worked and the epidemic is controlled, nevertheless it has been difficult to estimate the level of adherence to those policies and the repercussions in the sickness evolution curve. This effect can also be determinant during the present transition to the so-called 'new normal'.

The manuscript is organized as follows. In Section 2 we describe the construction of the social network from scratch using free data from Twitter, the opinion model is also introduced here and described its coupling to the epidemiological model. Section 3 contains the main findings and computations of the presented models, and Section 4 a summary and a brief discussion of the results, with conclusions and future perspectives.

## 2. Methods.

### 2.1 Social network construction

In order to generate a social network, we use Twitter. We downloaded several networks of connections (using the tool NodeXL [7]). Introducing a word of interest, NodeXL brings information of users that have twitted a message containing the typed word and the connections between them. The topics of the different searches are irrelevant. In fact, we tried to choose neutral topics with potentiality to engage many people independently of political commitment, age, or other distinctions. The importance of each subnet is that it reveals who is following who and allows us to build a more complete network of connections once all the subnets are put together. Each one of the downloaded networks will have approximately 2000 nodes [8]. In this way, downloading as many of such subnets as possible gives us a more realistic map of the current situation of the Spanish Twitter network and, we believe, a realistic approximation to the social interactions nationwide.

We intended to download diverse networks politically inoffensive. 'Junction' accounts will be needed to make sure that all sub-networks overlap. Junction accounts are these accounts that are part of several subnets and warrant the connection between them. If these junction accounts did not exist, isolated local small networks may appear. Go to supplementary information to see the word-of-interest networks downloaded and overlapped.

Twitter, as a social network, changes in time [9], [10], [11] and it is strongly affected by the current socio-political situation, so important variations in its configuration are expected with time. Specifically, when a major crisis, such as the current one, is ongoing. Taking this into consideration, we analyze two social neworks corresponding to different moments in time. One represents the social situation in October 2019 (with $N = 17665$ accounts) which describes a pre-epidemic social situation and another from April 2020 (with $N = 24337$ accounts) which describes the mandatory-confinement period of time. The networks obtained are directed and the links mark which nodes are following which. So, a node with high connectivity means it is following the opinions of many other nodes.

The two social networks obtained with this protocol are illustrated in figure 1. A first observation of their topologies demonstrate that they fit a scale free network with a power law connectivity distribution and exponents $\gamma = 1.39$ for October'19 and $\gamma = 1.77$ for April'20 network [12]. The significantly different exponents demonstrate the different internal dynamics of both networks.

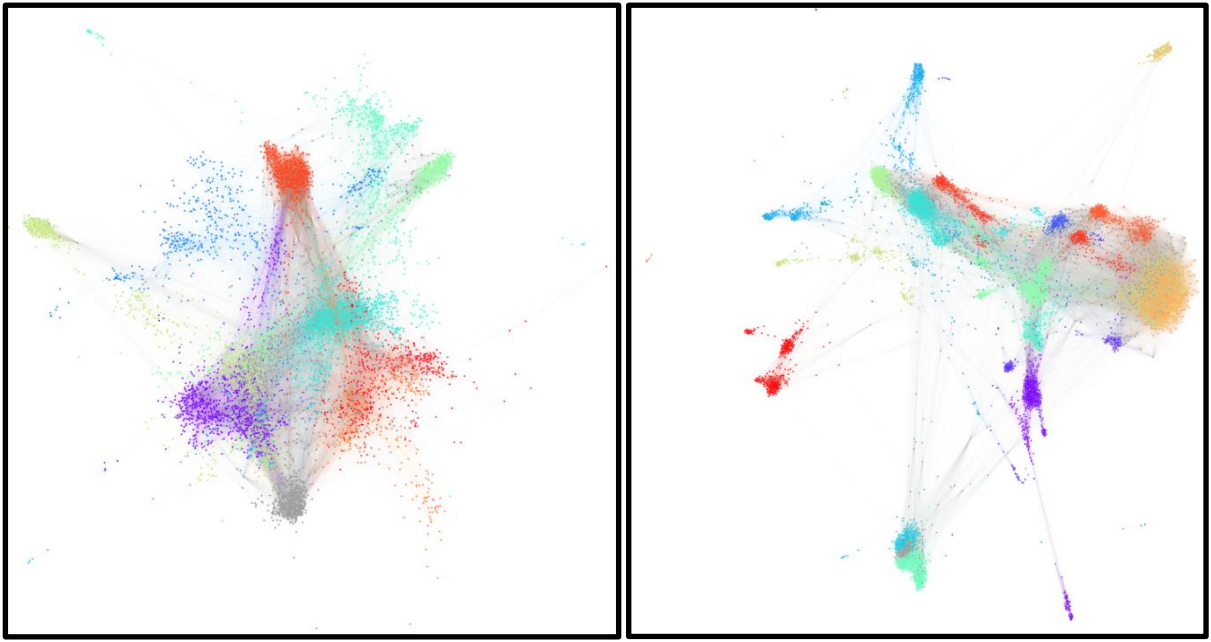

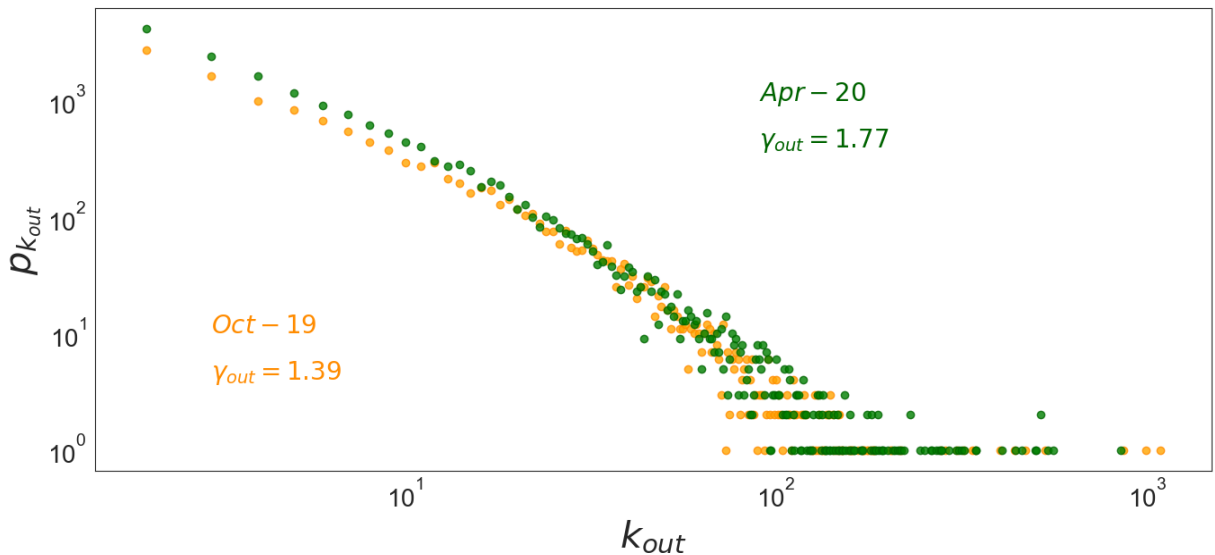

**Figure 1:** (a) October 2019 Twitter network. (b) April 2020 Twitter network. Each color marks those nodes corresponding with each word-of-interest subnet. Accounts in grey are the junction accounts. Links are colored with the origin node account. (c) Connectivity distribution for both networks. We generate the graphs in (a) and (b) using the algorithm Force Atlas 2 from Gephi [13]. Force atlas 2 is a forced-directed algorithm that stimulates the physical system to get the network organized through the space relying on a balance of forces; nodes repulse each other as charged particles while links attract their nodes obeying a Hooke's law. So that, nodes that are more distant exchange less information.

## 2.2 Opinion model

We consider a simple opinion model based on the logistic equation [14] but that has proved to be of use in other contexts [15, 16]. It is a two variable dynamical model whose nonlinearities are given by:

$$f(u,v) = Au\left(1 - \frac{u}{B}\right) + guv$$
$$g(u,v) = Cv\left(1 - \frac{v}{D}\right) - guv \tag{1}$$

where $u$ and $v$ account for the two different opinions. As $u + v$ remains constant, we can use the normalization equation $u + v = 1$, and, thus, the system reduces to a single equation:

$$f(u) = u\left[A\left(1 - \frac{u}{B}\right) + g(1 - u)\right] \tag{2}$$

$A$ is a time rate that modifies the rhythm of evolution of the variable $u$, $g$ is a coupling constant and $B$ controls the stationary value of $u$. This system has two fixed points ($u_0 = 0$ and $u_0 = \frac{A+g}{A/B+g}$ being the latest stable and $u_0 = 0$ unstable.

We now consider that each node belongs to a network and the connections between nodes follow the distribution measured in the previous section. The dynamic equation becomes [17],

$$\dot{u}_i = f(u_i) + d\frac{1}{k_i}\sum_{j=1}^{N} L_{ij}u_j \tag{3}$$

Each of the nodes $i$ obey the internal dynamic given by $f(u_i)$ while being coupled with the rest of the nodes with a strength $d/k_i$ where $d$ is a diffusive constant and $k_i$ is the connectivity degree for node $i$ (number of nodes each node is interacting with, also named outdegree). Note that this is a directed non-symmetrical network where $k_i$ means that node $i$ is following the Tweets from $k_i$ nodes. $L_{ij}$ is the Laplacian matrix, the operator for the diffusion in the discrete space, $i = 1, ..., N$. We can obtain the Laplacian matrix from the connections established within the network as $L_{ij} = A_{ij} - \delta_{ij}k_i$, being $A_{ij}$ the adjacency matrix

$$A_{i,j} = \begin{cases} 1 & \text{if } i,j \text{ are connected} \\ 0 & \text{if } i,j \text{ are not connected} \end{cases}$$

Notice that the mathematical definition in some references of the Laplacian matrix has the opposite sign. We use the above definition given by [17] in parallelism with Fick's law and in order to keep a positive sign in our diffusive system.

Now, we proceed as follows. We consider that all the accounts (nodes in our network) are in their stable fixed point $u_0 = \frac{A+g}{g+\frac{A}{B}}$, from equation (6), with a 10% of random noise. Then a subset of accounts $r$ is forced to acquire a different opinion, $u_i = 1$ with a 10% of random noise, $\forall i \ / \ i = 1,..rN$ and we let the system to evolve following the dynamical equations (3). In this case, accounts are sorted by the number of Followers that it is easily controllable. Therefore, some of the nodes shift their values to values closer to 1 that, in the context of this simplified opinion model, means that those nodes shifted their opinion to values closer to those leading the shift in opinion. This process is repeated in order to gain statistical significance and, as a result, it provides the probability distribution of nodes eager to change the opinion and adhere to the new politics.

## 2.3 Epidemiological model and coupling with opinion probability distribution

Our epidemiological model is based on the classic SIR model [18] and considers three different states for the population: susceptible (S), infected (I) and recovered or removed individuals (R) with the transitions as sketched in Figure 2.

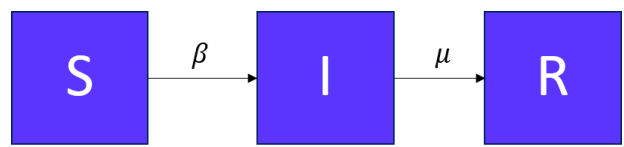

**Figure 2:** Scheme of the SIR model.

Here $\beta$ represents the probability of infection and $\mu$ the probability of recovering. We assume that recovered individuals gain immunity and therefore cannot be infected again. We consider an extended model to account for the epidemic propagation where each node interacts with others in order to spread the virus. In this context we consider that each node belongs to a complex network whose topology describes the physical interactions between individuals. The meaning of node here is a single person or a set of individuals acting as a close group (i.e. families). The idea is that the infected nodes can spread the disease with a chance $\beta$ to each of its connections with susceptible individuals, thus $\beta$ becomes a control parameter of how many individuals an infected one can propagate the disease to at each time step. Then, each infected individual has a chance $\mu$ of being recovered from the disease.

A first order approach to a human mobility network is the Watts-Strogatz model [19], given its ability to produce a clustered graph where nearest nodes have higher probability of being interconnected while keeping some chances of interacting with distant nodes (as in an Erdös-Renyi random graph [20]). According to this model, we generate a graph of $N$ nodes, where each node is initially connected to its $k$ nearest neighbors in a ring topology and the connections are then randomly rewired with distant nodes with a probability $p_{rewire}$. The closer this probability is to 1 the more resembling the graph is to a fully random network while for $p_{rewire} = 0$ it becomes a purely diffusive network. If we relate this ring-shaped network with a spatial distribution of individuals, when $p_{rewire}$ is small the occurrence of random interactions with individuals far from our circle of neighbors is highly severed, mimicking a situation with strict mobility restrictions where we are only allowed to interact with the individuals from our neighborhood. This feature makes the Watts-Strogatz model an even more suitable choice for the purposes of our study since it allows us to impose further mobility restrictions to our individuals in a simple way. On the other hand, the effects of clustering in small-world networks with epidemic models are important and have been already studied [21-24].

The network is initialized setting an initial number of nodes as infected while the rest are in the susceptible state and, then, the simulations starts. At each time step, the chance that each infected individual spreads the disease to each of its susceptible connections is evaluated by means of a Monte Carlo method [25]. Then, the chance of each infected individual being recovered is evaluated at the end of the time step in the same manner. This process is repeated until the pool of infected individuals has decreased to zero or a stopping criterion is achieved.

The following step in our modelling is to include the opinion model results from the previous section in the epidemic spread model just described. First, from the outcome of the opinion model $u$, we build a probability density $P(\bar{u})$ where $\bar{u} = 1 - u$ represents the disagreement with the externally given opinion. These opinion values are assigned to each of the nodes in the Watts-Strogatz network following the distribution $P(\bar{u})$. Next, we introduce a modified $\beta$ parameter, which varies depending on the opinion value of each node. It can be understood in terms of a weighted network modulated by the opinions, it is more likely that an infection occurs between two rogue individuals (higher value of $\bar{u}$) rather than between two individuals who agree with the government confinement policies ($\bar{u}$ almost zero or very close to zero). We introduce, then, the weight $\beta'_{ij} = \beta \cdot \bar{u}_i \cdot \bar{u}_j$, which accounts for the effective probability of infection between an infected node $i$ and a susceptible node $j$. At each time step of the simulation, the infection chances are evaluated accordingly to the value $\beta'_{ij}$ of the connection and the process is repeated until the pool of infected individuals has decreased to zero or the stopping criterion is achieved. In figure 3, we exemplify this process through a network diagram, where white, black and grey nodes represent susceptible, infected and recovered individuals respectively. Black connections account for possible infections with chance $\beta'_{ij}$.

To account for further complexity, this approach could be extrapolated to more complex epidemic models already presented in the literature [4, 6, 26]. Nevertheless, for the sake of illustration, this model still preserves the main features of an epidemic spread without adding the additional complexity to account for real situations such as the COVID19 case.

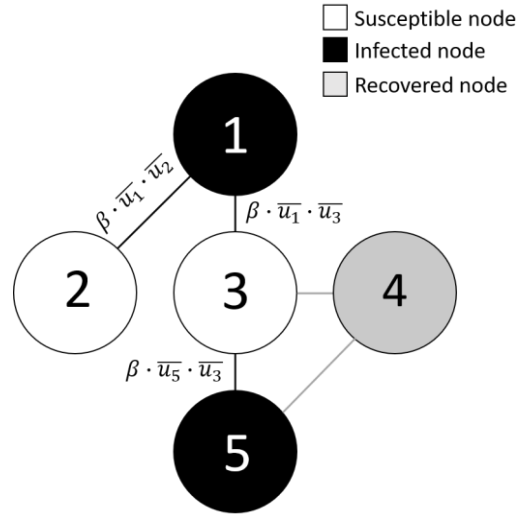

**Figure 3.** Diagram of the infection process in the network. Black links account for possible infections with weight $\beta'_{ij} = \beta \cdot \bar{u}_i \cdot \bar{u}_j$.

## 3. Results

### 3.1 Social Network

Following the previous protocol, we run the opinion model considering the two social networks analyzed. Figure 4 shows the distribution of the final states of the $u$ variable for the October'19 network (orange) and the April'20 network (green) when the new opinion is introduced in a 30% of the total population (r=30%). Different percentages of the initial population *r* were considered but the results are equivalent (see figure S1 in the supplementary information).

Direct inspection of Figure 4 clearly shows that the population on April'20 is more eager to follow the new opinion (political guidelines) comparing with the situation in October'19. In the pandemic scenario (network of April'20) it is noticeable that larger values of the opinion variable, $u_i$, are achieved corresponding with the period of the quarantine. Preferential states are also observed around $u_i = 0$, $u_i = 0.5$ and $u_i = 1$. Note that the network of April'20 allows to change opinions more easily than in the case of October'19.

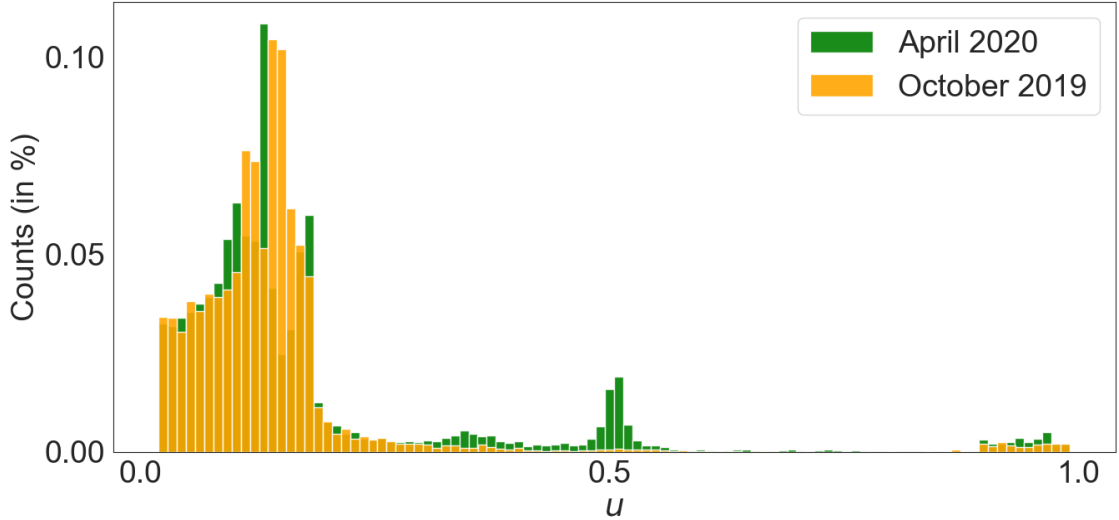

**Figure 4.** Distribution of the concentrations $u_i$ for the Twitter network from October 2019 (orange) and April 2020 (green) for a r=30% of the initial accounts in the state 1 with a 10% of noise ($A$=0.0001, $B$=0.01, $g$=0.0001, $x0$=0.01, $d$=20000).

### 3.2 Opinion biased epidemic model

During the sanitary crisis in Spain, the government imposed heavy restrictions on the mobility of the population. To better account for this situation, we rescaled the probability density of disagreement opinions $P(\bar{u})$ to values between 0 and 0.3, leading to the probability densities of figure 5. From here on, we shall refer to this maximum value of the rescaled probability density as the ***cutoff*** imposed to the probability density. Note that this probability distribution is directly included into de mobility model as a probability to interact with other individuals, thus, this cutoff means that the government policy is enforced reducing up to a 70% of the interactions and the reminder 30% is controlled by the population decision to adhere to the official opinion.

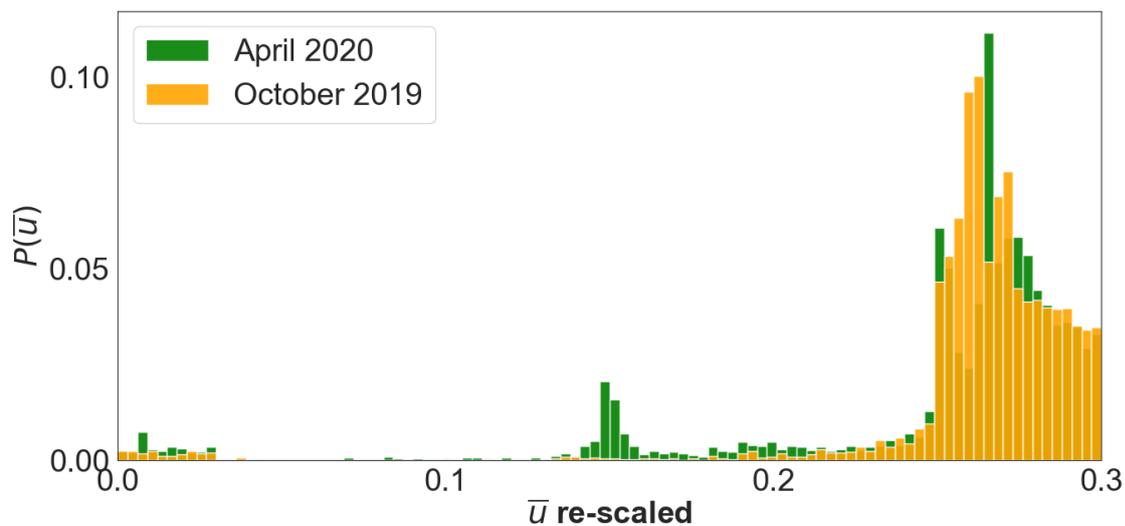

**Figure 5**. Probability densities of the variable $\bar{u} = 1 - u$ constructed from the distributions of figure 4 and rescaled to the values between 0 and 0.3 to account for a heavily restricted mobility.

In figure 6 we summarized the main results obtained from the incorporation of the opinion model into the epidemiological one. We established four different scenarios: for the first one we considered a theoretical situation where we imposed that around the 70% of the population will adopt social distancing measures, but leave the other 30% in a situation where they either have an opinion against the policies or they have to move around interacting with the rest of the network for any reason (this means, $\bar{u} = 0.3$ for all the nodes). In contrast to this situation we introduce the opinion distribution of the social networks of April'20 and October'19. Finally, we consider another theoretical population where at least 90% of the population will adopt social distancing measures (note that in a real situation, around 10% of the population occupies essential jobs and, thus, are still exposed to the virus). However, for the latter the outbreak of the epidemic does not occur so there is no peak of infection. Note that the first and the last ones are completely in-silico scenarios introduced for the sake of comparison.

Figure 6a shows the temporal evolution of the infected population in the first three of the above scenarios. The line in blue shows the results without including an opinion model and considering that a 70% of the population blindly follows the government mobility restrictions while the reminding 30% continue interacting as usual. Orange line shows the evolution including the opinion model with the probability distribution derived as in October'19. The green line is the evolution of the infected population considering the opinion model derived from the situation in April'20. Note that the opinion model stated that the population in April'20 was more eager to follow changes in the opinion than in October'19, and this is directly reflected in the curves in Figure 6a. Also note that as the population becomes more conscious and decides to adhere to the restriction-of-mobility policies, the maximum of the infection curve differs in time and its intensity is diminished. This figure clearly shows that the state of the opinion inferred from the social network analysis strongly influences the evolution of the epidemic.

The results from the first theoretical case (blue curve) show clearly that the disease reaches practically all the rogue individuals (around the 30% of the total population that we set with the rescaling of the probability density), while the other two cases with real data show that further agreement with the given opinion results in flatter curves of infection spreading. We have analyzed both the total number of infected individuals on the peaks and its location in time of the simulation, but, since our aim is to highlight the incorporation of the opinion model we show in Figures 6b and 6c the values of the maximum peak infection as well as the delay introduced in achieving this maximum scaled with the corresponding values of the first case (blue line). We see that the difference on the degree of adhesion of the social networks outcomes a further 12% reduction approx. on the number of infected individuals at the peak, and a further delay of around the 20% in the time at which this peak takes place. Note that for the April'20 social network, a reduction of almost the 50% of individuals is obtained for the peak of infection, and a similar value is achieved for the time delay of the peak. This clearly reflects the fact that a higher degree of adhesion is important to flatten the infection curve. Finally, in the latter theoretical scenario, where we impose a cutoff of $\bar{u} = 0.1$, the outbreak of the epidemic does not occur, and thus there is no peak of infection. This is represented in figure 6b and 6c as a dash-filled bar indicating the absence of the said peak.

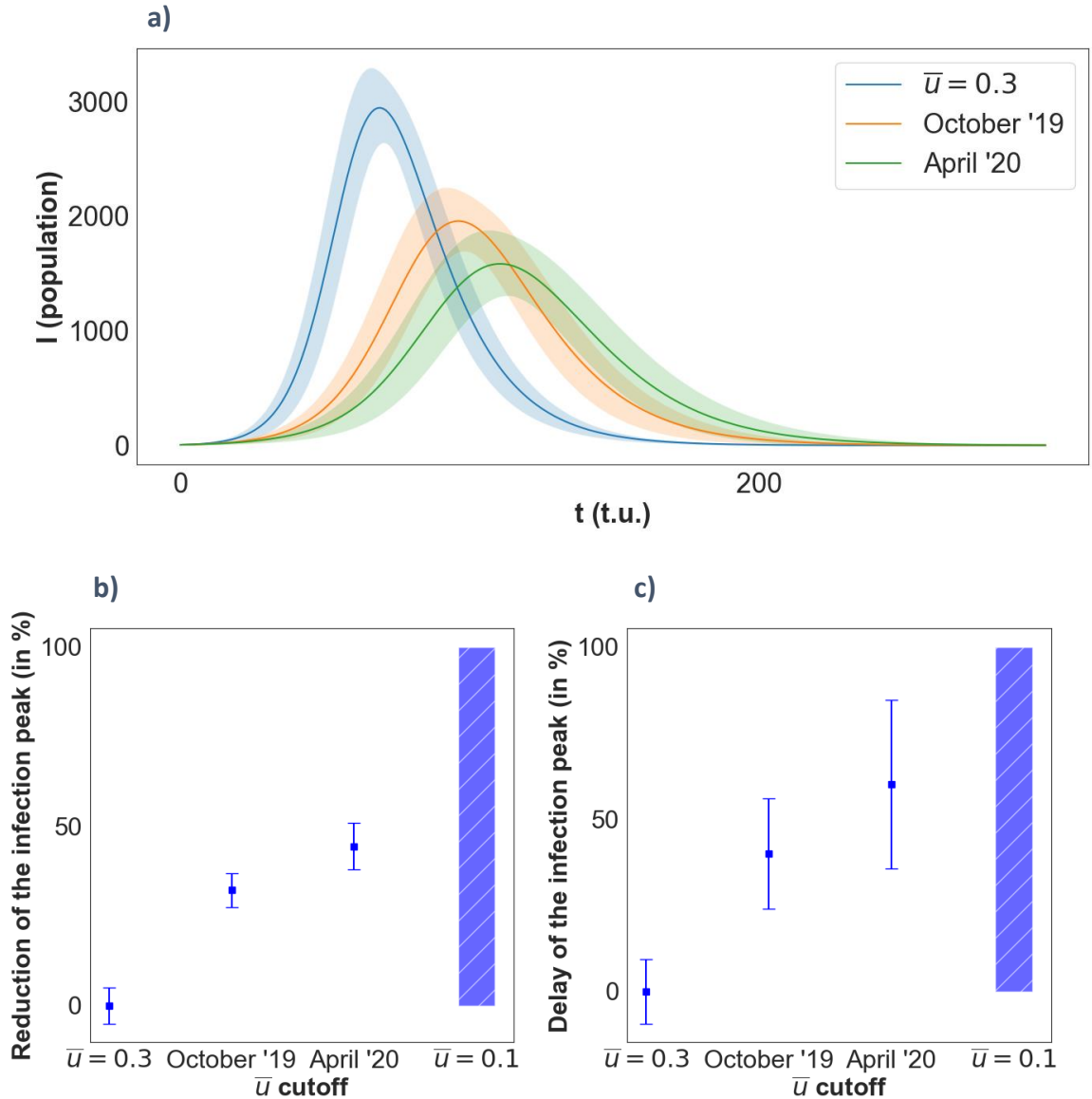

**Figure 6.** (a) Evolution of the number of infected individuals with time for the three opinion models considered. (b) Reduction in % of the infected individuals on the peak of the curve in respect to the model with fixed opinion $\bar{u} = 0.3$. (c) Time delay in % of the infection peak. Error bars represent the standard deviation of a sample of peak statistics obtained across several simulations with the same parameters, but different random configurations of the network's adjacency matrix. ($N = 10000$, $\beta = 0.05$, $\mu = 0.06$, $p_{rewire} = 0.25$).

Changing the condition on the cutoff imposed for the variable $\bar{u}$ can be of interest to model milder or stronger confinement scenarios such as the different policies ruled in different countries. In figure 7 we show the infection peak statistics (maximum of the infection curve and time at maximum) for different values of the cutoffs and for both social opinion networks. In both cases, the values are scaled with those from the theoretical scenario with all individuals having their opinion at the cutoff value. Both measurements (Figures 7a and 7b) are inversely proportional to the value of the cutoff. This effect can be understood in terms of the obtained probability densities. For both networks (October'19 and April'20) we obtained that most of the nodes barely changed their opinion, and thus for increasing levels on the cutoff of $\bar{u}$ these counts dominate on the infection processes so the difference between both networks is reduced. On the other hand, this highlights the importance of rogue individuals in situations with increasing levels of confinement policies since for highly contagious diseases each infected individual propagates the disease rapidly. Each infected individual matter and the less connections he or she has the harder is for the virus to spread along the exposed individuals. Note that for all the scenarios, the social network of April'20 represents the optimum situation in terms of infection peak reduction and its time delay. It is particularly interesting the case for the cutoff in $\bar{u} = 0.2$. All simulations run for this cutoff show an almost non-existent peak. This is represented on figure 7a with almost a reduction of the 100% of the infection peak (the maximum value found on the infection curve was small but not zero) and the value of the time delay (Figure 7b) is included in the shaded region since this infection curve was almost flat. Something similar occurs for the cutoff in $\bar{u} = 0.25$, which explains the large error seen in figure 7b. Note that this value of the cutoff, $\bar{u} = 0.2$, constitutes by itself an actual threshold bellow which no infection peak is observed.

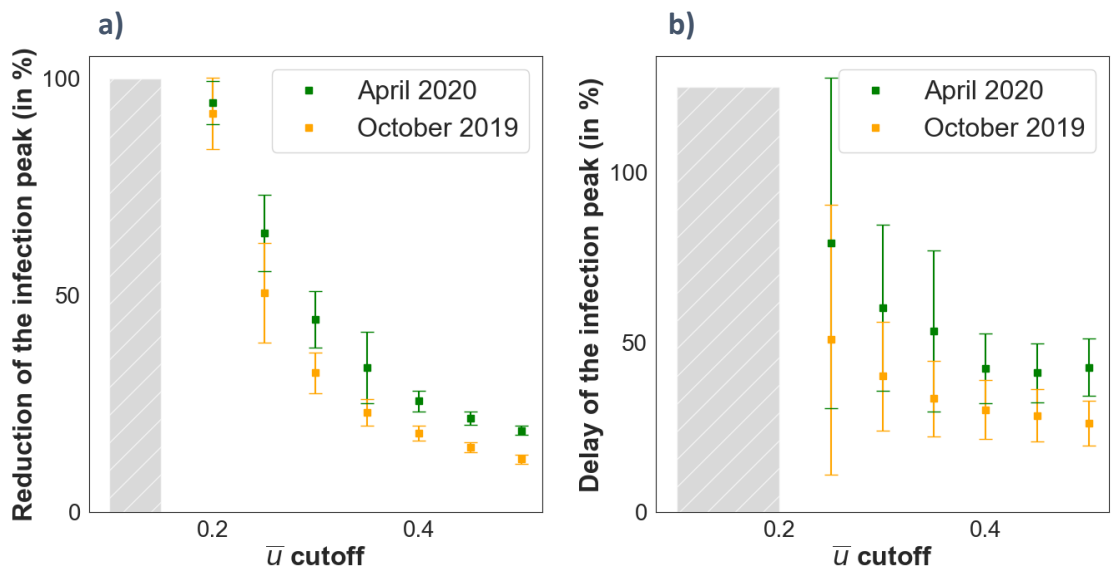

**Figure 7.** Infection peak statistics for different values of the cutoff of $\bar{u}$ and for both the opinion models (October'19 in yellow and April'20 in green). (a) Reduction of the maximum in the infection curve scaled with the corresponding maximum of the least favorable case (theoretical scenario where all nodes have the opinion of the cutoff). (b) Delay of the maximum in the infection curve scaled with the corresponding time for the maximum of the least favorable case. Again, the dash-filled bar represents the absence of an infection peak, and the error bars represent the standard deviation of a sample of peak statistics obtained across several simulations. See figure S3 on the supplementary information for the time evolution of the infected individuals of some of the points shown here. *(N = 10000, $\beta$ = 0.05, $\mu$ = 0.06, $p_{rewire}$ = 0.25).*

As discussed in the previous section, we are considering a Watts-Strogatz model for the mobility network. This type of network is characterized by a probability of rewiring (as introduced in the previous section) that stablishes the number of distant connections for each individual in the network. All previous results were obtained considering a probability of rewiring of 0.25. Figure 8 shows the variation of the maximum for the infection curve and time for the maximum versus this parameter. The observed trend indicates that the higher the clustering (thus, the lower the probability of rewiring) the more difficult is for the disease to spread along the network. This result is supported by previous studies in the field, which show that clustering decreases the size of the epidemics and in cases of extremely high clustering, it can die out within the clusters of population [21,24]. This can be understood in terms of the average shortest path of the network [12], which is a measure of the network topology that tells the average minimum number of steps required to travel between any two nodes of the network. Starting from the ring topology, where only the nearest neighbors are connected, the average shortest path between any two opposite nodes is dramatically reduced with the random rewirings. Remember that these new links can be understood as short-cuts or long-distance connections within the network. Since the infection process can only occur between active links between the nodes, it makes sense that the propagation is limited if less of these long-distance connections exist in the network. The average shortest path length decays extremely fast with increasing values of the random rewiring, and thus we see that the peak statistics are barely affected for random rewirings larger than the 25%. If one is interested on further control of the disease, the connections with distant parts of the network must be minimized to values smaller than this fraction. Regarding the performance of both opinion biased epidemic cases, we found again a clear difference between the two of them. In the April'19 case, the outcome of the model present always a more favorable situation to control the expansion of the epidemic, stating the importance of the personal adherence to isolation policies in controlling the evolution of the epidemic.

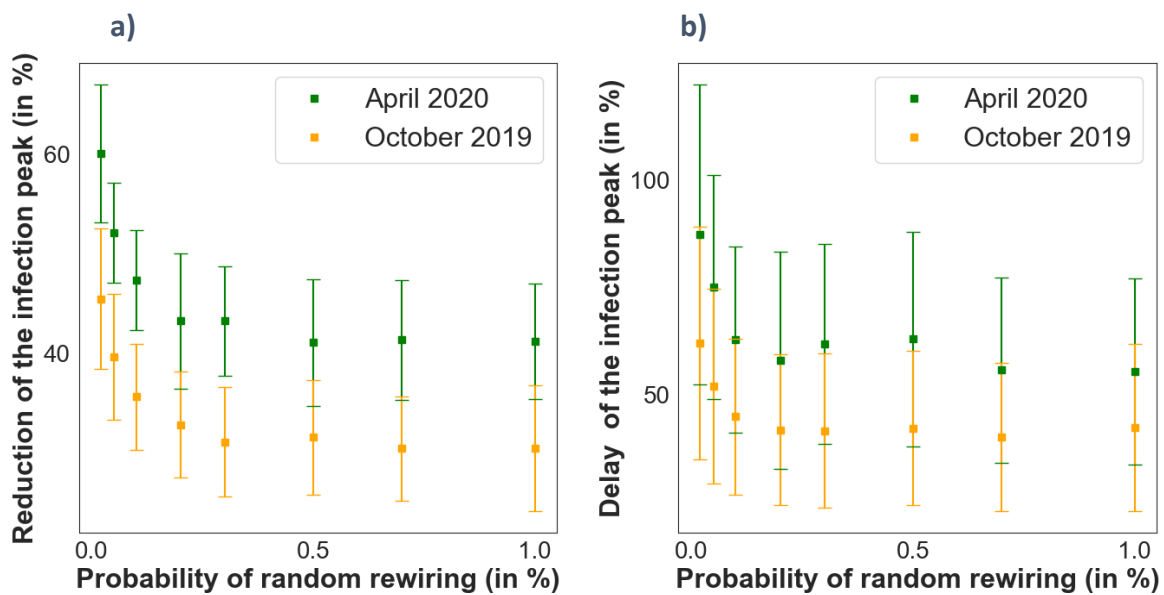

**Figure 8.** Peak statistics of the infection curves for different values of the rewiring probability of the Watts-Strogatz model, and for both opinion scenarios October'19 and April'20. See figure S4 on the supplementary information for the time evolution of the infected individuals of some of the points shown here. ($N = 10000$, $\beta = 0.05$, $\mu = 0.06$, $\bar{u}$ cutoff $= 0.3$)

## 4. Discussion

We have parametrized the social situation of the Spanish society at two different times with the data collected from a social media based on microblogging (twitter.com). The topology of these networks combined with a simple opinion model provides us with an estimate of how likely this society is to follow new opinions and change their behavioral habits. The first analysis presented here shows that the social situation in October 2019 differs significantly from that of April 2020. In fact, we have found that the latter is more likely to accept opinions or directions and, thus, follow government policies such as social distancing or confining. The output of these opinion models was used to tune the mobility in an epidemic model aiming to highlight the effect that the social 'mood' has on the pandemic evolution. The histogram of opinions was directly translated into a probability density of people choosing to follow or not the directions, modifying their exposedness to being infected by the virus. Although we exemplify the results with an over-simplified epidemic model (SIR), the same protocol can be implemented in more complicated epidemic models. We show that the partial consensus of the social network, although non perfect, induces a significant impact on the infection curve, and that this impact is quantitatively stronger in the network of April 2020. Our results are susceptible to be included in more sophisticated models used to study the evolution of the COVID19.

All epidemic models lack to include the accurate effect of the society and their opinions in the propagation of epidemics. We propose here a way to monitor, almost in real time, the mood of the society and, therefore, include it in a dynamic epidemic model that is biased by the population eagerness to follow the government policies.

Further analysis of the topology of the social network may also provide insights of how likely the network can be influenced and identify the critical nodes responsible for the collective behavior of the network.

## Acknowledgments

This research is supported by the Spanish Ministerio de Economía y Competitividad and European Regional Development Fund, research grant No. COV20/00617 and RTI2018-097063-B-I00 AEI/FEDER, UE; by Xunta de Galicia, Research Grant No. 2018-PG082, and the CRETUS Strategic Partnership, AGRUP2015/02, supported by Xunta de Galicia. All these programs are co-funded by FEDER (UE). We also acknowledge support from the Portuguese Foundation for Science and Technology (FCT) within the Project n. 147.

# Supplementary information

# Incorporating social opinion in the evolution of an epidemic spread

Alejandro Carballosa, Mariamo Mussa Juane and Alberto P. Muñuzuri

Institute CRETUS. Group of Nonlinear Physics. Fac. Physics. University of Santiago de Compostela. 15782 Santiago de Compostela, Spain

**Supplementary contents**

1. List of Hashtags used to build up the social networks considered.

2. Opinion distributions depending on the initial number of nodes with different opinion.

3. Opinion biased epidemic model

# 1. List of Hashtags used to build up the social networks considered.

The list of hashtags used to construct both networks is in Table 1 for the October'19 case (column on the left) and for the April'20 scenario (right column). All hashtags used were neutral in the sense of political bias or age meaning.

| October'19 | April'20 |
| --- | --- |
| #eleccionesgenerales28a | #CuidaAQuienTeCuida |
| #eldebatedecisivolasexta | #EsteVirusLoParamosUnidos |
| #PactosARV | #QuedateConESP |
| #RolandGarros | #SemanaEnCasaYoigo |
| #NiUnaMenos | #QuedateEnCasa |
| #selectividad2019 | #Superviviente2020 |
| #AnuncioEleccions28Abril | #AutonomosAbandonados |
| #BlindarElPlaneta | #Renta2019 |
| #DiaMundialDeLaBicicleta' | #EnCasaConSalvame |
| #EmergenciaClimatica27S' | #diamundialdelasalud |
|  | #CuarentenaExtendida |
|  | #AsiNonUvigo |
|  | #AhoraTocaLucharJuntos |
|  | #House_Party |
|  | #EnCasaConSalvame |
|  | Apoyare_a_Sanchez |
|  | Pleno_del_Congreso |
|  | Viernes_de_Dolores |

**Table 1:** List of hashtags used to construct the networks.

In order to check the statistical accuracy and relevance of our networks, we considered different scenarios with more or less subnets (each subnet corresponding with a single hashtag) and estimate the exponent of the scale-free-network fit. This result is illustrated in Figure S1a for the October'19 case and in Figure S1b for the April'20 case. Note that as the number of subnets (hashtags) is increased, the exponent converges.

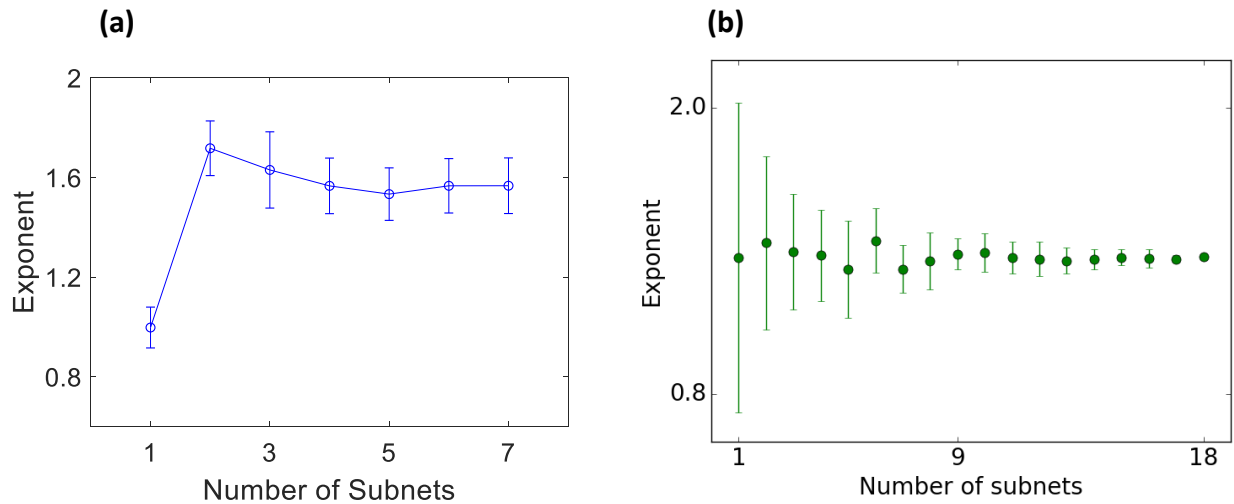

**Figure S1.** Variation of the exponent versus the number of subnets considered (a) October 2019. (b) April 2020 exponent of the scale free distribution. Each one of the exponents was calculated merging 10 combinations of $2, 3, …, N-1$ subnets. The error bars are the standard deviation. For 1 subnet all the exponents were calculated and for N subnets just one combination is possible so that non deviation is shown.

**2. Opinion distributions depending on the initial number of nodes with different opinion.**

Distribution of the final states of the $u$ variable for the October'19 network (orange) and the April'20 network (green) when the new opinion is introduced by three different percentages of the total population (r parameter) is shown in figure S2. Note that in all cases the results are qualitatively equivalent and, once included in the opinion model, the results are similar.

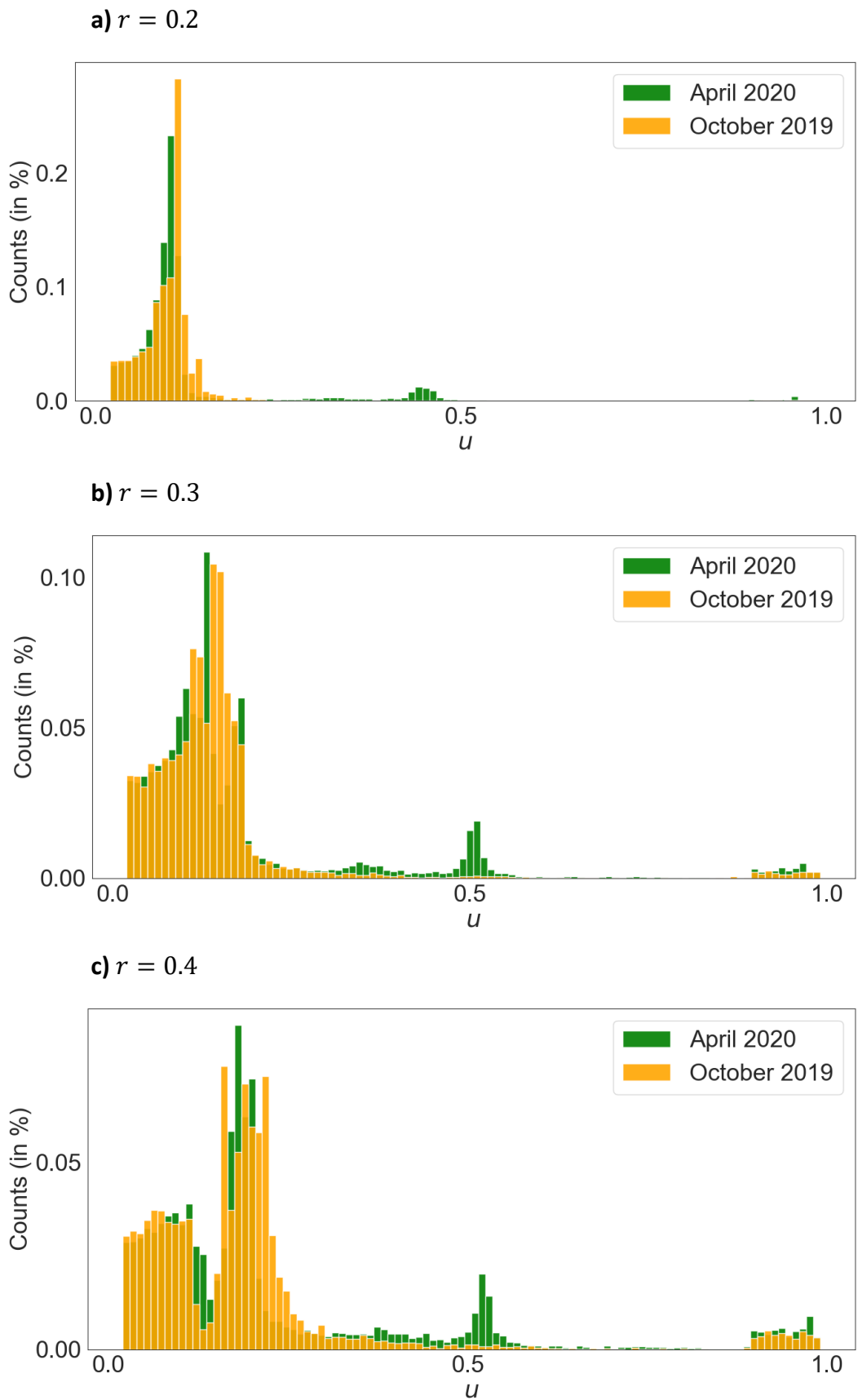

**Figure S2.** Distribution of the concentrations $u_i$ for the Twitter network from October 2019 (orange) and April 2020 (green) for r=20% (a), r=30% (b) and r=40% (c) of the initial accounts in the state 1 with a 10% of noise ($A$=0.0001, $B$=0.01, $g$=0.0001, $x0$=0.01, $d$=20000).

## 3. Opinion biased epidemic model

Figure S3 shows the evolution of the number of infected individuals with time for the epidemic model biased with the opinion model of April 2020. Results for different values of the $\bar{u}$ cutoff are shown. Note how for $\bar{u} = 0.2$ the peak of infection vanishes, and the epidemic dies out due to its lack of ability to spread among the nodes. On the other hand, Figure S4 shows for different values of the cutoff on $\bar{u}$, the comparison between the three cases presented in the main text (see figure 6): the theoretical scenario where the opinion is fixed on the cutoff value for all the nodes, and the epidemic model biased with the opinions of October '19 and April '20 scenarios. See how the difference between the theoretical scenario and the opinion biased models diminishes with growing values of the cutoff value on $\bar{u}$

Finally, Figure S5 shows the effect that higher values of the rewiring probability of the Watt-Strogatz model has in the time evolution of the infected individuals. As shown in the main text, lower values of the rewiring probability has an important impact on the peak of infection, while values above $p_{rewiring} = 0.3$ barely change the statistics on the said peak, or fall within the error of the measurements.

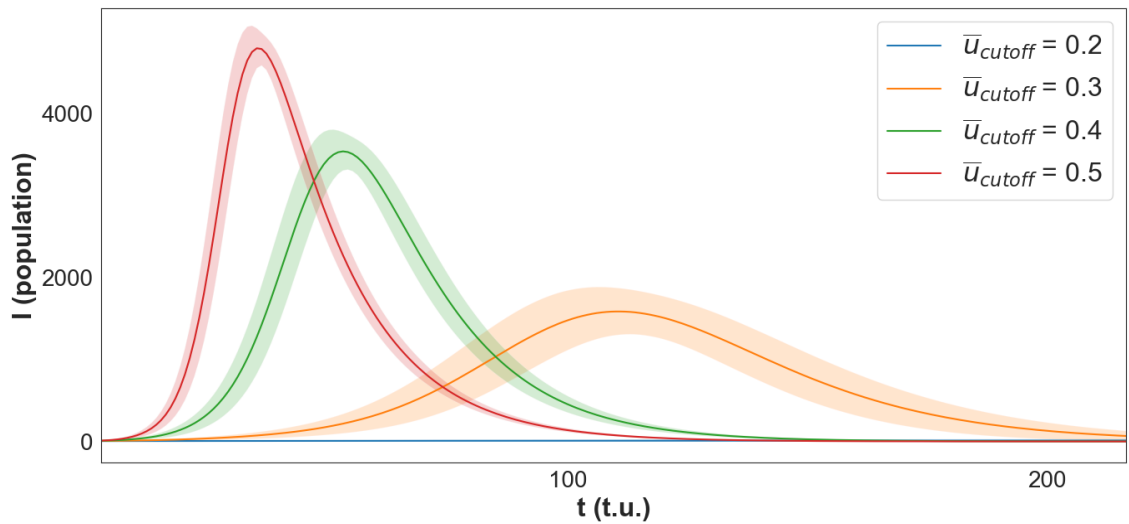

**Figure S3.** Evolution of the number of infected individuals with time for the epidemic model biased with the April'20 social network and for different values of the cutoff on $\bar{u}$. ($N = 10000$, $\beta = 0.05$, $\mu = 0.06$, $p_{rewire} = 0.25$)

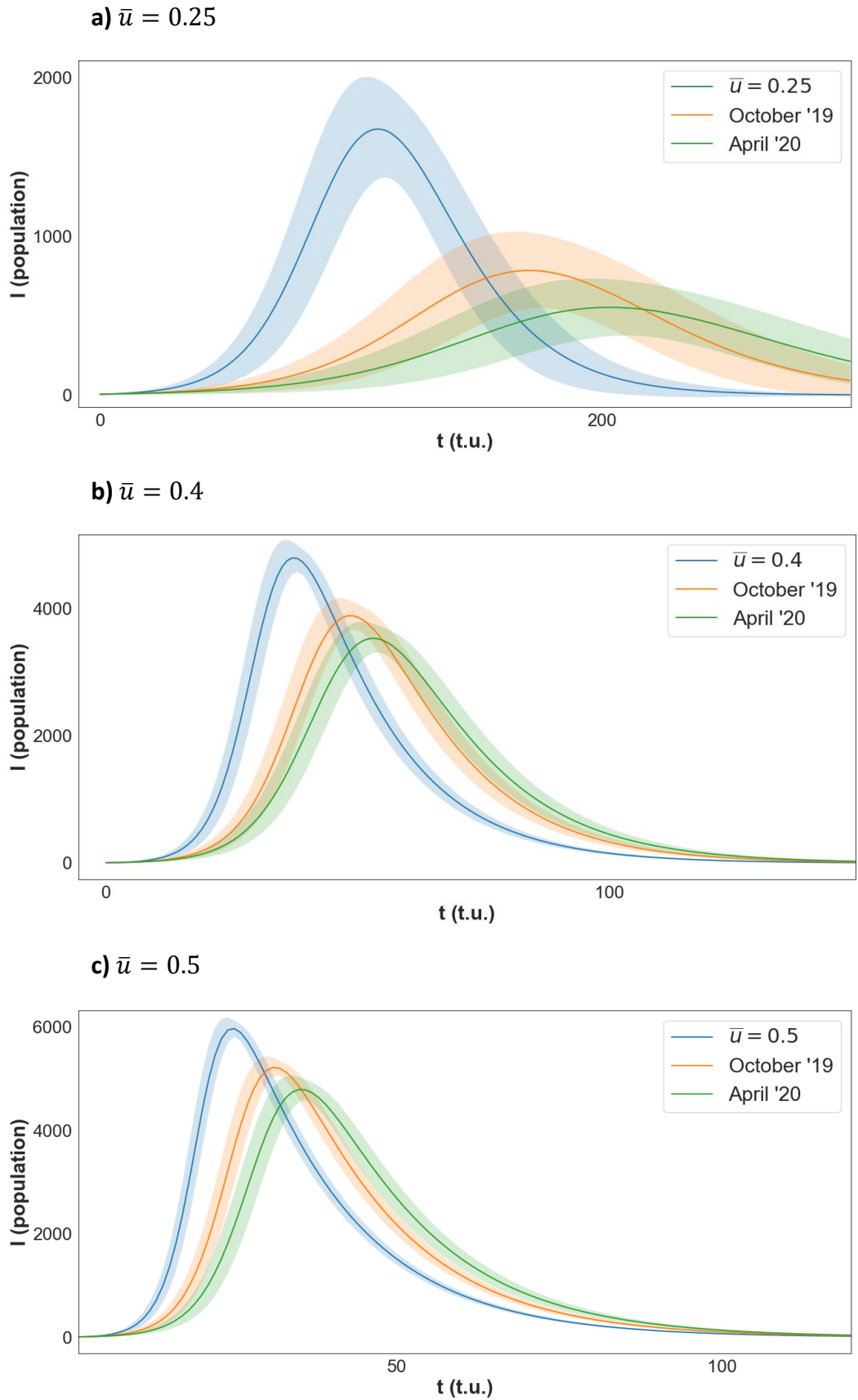

**Figure S4.** Evolution of the number of infected individuals with time for the three opinion models considered for three different values of the cutoff on $\bar{u}$: a) $\bar{u}$=0.25, b) $\bar{u}$=0.4 and c) $\bar{u}$=0.5. ($N = 10000$, $\beta = 0.05$, $\mu = 0.06$, $p_{rewire} = 0.25$)

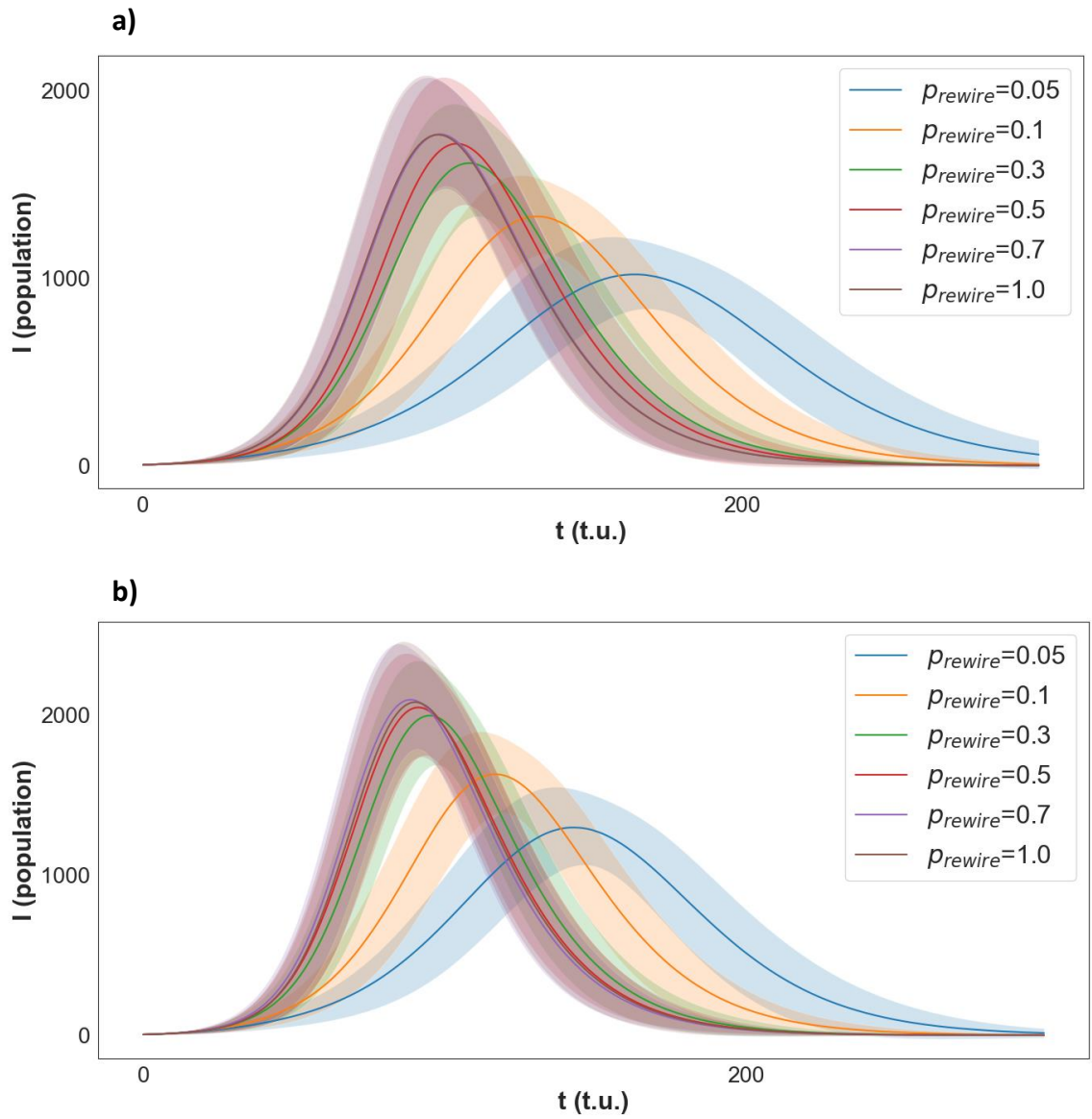

**Figure S5.** Evolution of the number of infected individuals with time for the epidemic models biased with the April'20 social network (a) and the October'19 social network (b), for different values of the rewiring probability of the Watt-Strogatz network model. ($N = 10000$, $\beta = 0.05$, $\mu = 0.06$, $\bar{u}\ cutoff = 0.3$)